\documentclass[prl,twocolumn,floatfix,superscriptaddress]{revtex4-1}
\usepackage{dcolumn,amsmath}
\usepackage{graphicx}
\usepackage{bm}
\usepackage{hyperref} 
\newcommand{\Eeff}{\ensuremath{E_{\rm eff}}}
\newcommand{\eEDM}{{\em e}EDM}
\newcommand{\ecm}{\ensuremath{e {\cdotp} {\rm cm}}}

\newcommand{\de}{d_\mathrm{e}}
\begin{document}
   \title{Theoretical study of ThF$^+$ in the search for T,P-violation effects:
Effective state of a Th atom in ThF$^+$ and ThO compounds}

\author{L.V.\ Skripnikov}\email{leonidos239@gmail.com}
\author{A.V.\ Titov}
\homepage{http://www.qchem.pnpi.spb.ru}
\affiliation{National Research Centre ``Kurchatov Institute'' B.P. Konstantinov Petersburg Nuclear Physics Institute, Gatchina, Leningrad district 188300, Russia}
\affiliation{Dept.\ of Theoretical Physics, St.Petersburg State University, 198504, Russia}

%\author{???}
%\affiliation{???}

%\date{\today, 18la}
\date{01.03.2015}

\begin{abstract}

We report the results of theoretical investigation of electronic structure of ThF$^+$\ 
cation which is one of the most interesting systems to search for the permanent electric dipole moment (EDM) of the electron (\eEDM) [H.~Loh, K.C.~Cossel, M.C.~Grau, K.-K.~Ni, E.R.~Meyer, J.L.~Bohn, J.~Ye, E.A.~Cornell, Science {\bf 342}, 1220 (2013)] and other effects of violation of time reversal (T) and spacial parity (P) symmetries in fundamental interactions. For the working $^3\Delta_1$ state we have found a quite high value of the effective electric field acting on unpaired electrons (37.3 GV/cm). The field will be required to interpret the experiment planed on ThF$^+$ in terms of \eEDM. Within the concept of atoms in compounds [A.V.~Titov, Y.V.~Lomachuk, and L.V.~Skripnikov, Phys.\ Rev.~A {\bf 90}, 052522 (2014)] we have compared the ThF$^+$ electronic structure with that of ThO. Also we have calculated other parameters of T,P-odd interactions: $W_{T,P}$, which is needed for interpretation of the experiment in terms of the dimensionless constant $k_{T,P}$ characterizing the strength of the T,P-odd pseudoscalar$-$scalar electron$-$nucleus neutral current interaction (50~kHz); $W_M$, which is required to search for the Th nuclear magnetic quadrupole moment in $^{229}$ThF$^+$ (0.88 $\frac{10^{33}\mathrm{Hz}}{e~{\rm cm}^2}$).
A number of properties which can be measured are also calculated: hyperfine structure constant, the molecule-frame dipole moment, and g-factor.
\end{abstract}

\maketitle

%#####################
\section{Introduction}
%#####################

During the last decade an impressive progress in the search for the permanent electric dipole moment of the electron (\eEDM) has been achieved \cite{Regan:02, Hudson:11a, ACME:14a}. The great interest in \eEDM\ is caused by the fact that its nonzero value implies manifestation of interactions which are not symmetric with respect to both time (T) and space (P) inversions (T,P-odd interactions). According to the Standard model \eEDM\ should be less than $10^{-38} \ecm$ \cite{Khriplovich:11}. Therefore the observation of \eEDM\ at a notably higher level would indicate the presence of a ``new physics'' beyond the Standard model. Most popular extensions of the Standard model predict the magnitude of the \eEDM\ at the level of $10^{-26}-10^{-29} \ecm$ \cite{Commins:98} and that range is almost passed to-date by the latest atomic and molecular measurements.

It was found since sixties of the past century \cite{Sandars:64, Sandars:67, Shapiro:1968, Sushkov:84, Labzowsky:78, Ginges:04, Khriplovich:11,Chubukov:15} that extremely sensitive experiments towards the search of T,P-odd effects can be performed on heavy-atom molecules and solids. The current limit, $|\de|<8.7\times 10^{-29}$\ \ecm\ (90\% confidence), was set with a molecular beam of thorium monoxide (ThO) molecules in the metastable electronic $H^3\Delta_1$ state \cite{ACME:14a}. The previous best limit was also established on a molecular beam but using the YbF radicals \cite{Hudson:11a}.

Nowadays, a number of new prospective systems are suggested, investigated theoretically and, in part, prepared experimentally (HfF$^+$ \cite{Cossel:12, Cornell:13, Petrov:07a, Fleig:13, Meyer:06a}, YbF \cite{Hudson:11a, Mosyagin:98, Quiney:98, Parpia:98, Kozlov:97c, Nayak:09, Steimle:07, Abe:14}, ThO \cite{ACME:14a, Petrov:14, Meyer:08, Skripnikov:13c, Skripnikov:14a,Skripnikov:14b, Fleig:14}, ThF$^+$ \cite{Cornell:13}, WC \cite{Lee:13a, Meyer:09a}, PbF \cite{Shafer-Ray:08E, Skripnikov:14c, Petrov:13},RaO \cite{Flambaum:08,Kudashov:13}, RaF \cite{Isaev:12,Kudashov:14} etc.) which promise to achieve a sensitivity to \eEDM\ up to $10^{-29}-10^{-30}\ecm$. One of promising experiments towards the measurement of \eEDM\ is proposed on the $^3\Delta_1$ state of the cation of thorium monofluoride (ThF$^+$) by E.~Cornell group \cite{Cornell:13}. The use of the $^3\Delta_1$ state has a number of advantages from experimental point of view. Due to $\Omega$-doublet structure of $^3\Delta_1$ state the interval between the opposite parity levels is very small. Therefore, the molecule can be polarized by a weak electric field which leads to cancellation of some systematic errors since the effect on the doublet components has an opposite sign \cite{SF78,ComminsFest, Petrov:14}. Also, magnetic moment (g-factor) of the $^3\Delta_1$ electron state is very small (zero in the  nonrelativistic limit), and this is another reason for reducing the systematic errors. The advantage of using such a state has been demonstrated in the recent experiment on ThO molecule \cite{ACME:14a}.

The working $^3\Delta_1$ state of ThO is a metastable (first excited) one with the lifetime of about 2~ms \cite{Vutha:2010} whereas the ground state is $^1\Sigma$. In contrast to the ThO case, the energies of the $^3\Delta_1$ and $^1\Sigma$ states in ThF$^+$ are very close \cite{Barker:2012}. 
%new
%Moreover, it was found only recently that the $^3\Delta_1$ state in ThF$^+$ is also the first excited state \cite{Heaven:14}. However, the transition energy is far smaller than that in ThO (316 \cm\ vs.\ 5321 \cm\ in ThO \cite{Huber:79, Edvinsson:84}). This suggests a very big lifetime of the working state and a good statistics.
In Ref. \cite{Heaven:14} the $^3\Delta_1$ state of ThF$^+$ was assigned as the first excited state with transition energy 
316 cm$^{-1}$ (versus 5321 cm$^{-1}$ in ThO \cite{Huber:79}). However, the most recent experiments by Cornell group show that $^3\Delta_1$ is the ground state of ThF$^+$ \cite{Cornell:15}.
This suggests a very good statistics.
%end new

%======================
%Effective Hamiltonian.
%======================

To interpret the results of the ThF$^+$ experiment in terms of the \eEDM\ one should know a parameter usually called ``the effective electric field on electron'', \Eeff,
which cannot be measured. \Eeff\ actually relevant to only the spin-polarized electrons (the closed shells do not contribute to measured effects in context of \eEDM, see next section), it can be evaluated as an expectation value of the T,P-odd operator (see Refs.\ \cite{Kozlov:87, Kozlov:95, Titov:06amin}):
\begin{equation}
W_d = \frac{1}{\Omega}
\langle \Psi|\sum_i\frac{H_d(i)}{d_e}|\Psi\rangle,
\label{matrelem}
\end{equation}
where $d_e$ is the value of \eEDM,$\Psi$ is the wave function of the considered state of ThF$^+$, and $\Omega= \langle\Psi|\bm{J}\cdot\bm{n}|\Psi\rangle$,
$\bm{J}$ is the total electronic momentum, $\bm{n}$ is the unit vector along the molecular axis $\zeta$ directed from Th to F ($\Omega=1$ for the considered $^3\Delta_1$ state of ThF$^+$),
\begin{eqnarray}
  H_d=2d_e
  \left(\begin{array}{cc}
  0 & 0 \\
  0 & \bm{\sigma E} \\
  \end{array}\right)\ ,
 \label{Wd}
\end{eqnarray}
$\bm{E}$ is the inner molecular electric field, and $\bm{\sigma}$ are the Pauli matrices. In these designations $E_{\rm eff}=W_d|\Omega|$.

Besides the interaction given by operator (\ref{Wd}) there is a T,P-odd pseudoscalar$-$scalar electron$-$nucleus neutral currents interaction with the dimensionless constant $k_{T,P}$. Note that it was estimated in Ref.~\cite{Pospelov:14} within the Standard model that this interaction can induce even greater T,P-odd effect in ThO simulating the \eEDM. The interaction is given by the following operator~\cite{Hunter:91}:
\begin{eqnarray}
%  H_{T,P}=i\frac{G\alpha}{\sqrt{2}}Zk_{T,P}\gamma_0\gamma_5n(\textbf{r}),
H_{T,P}=i\frac{G_F}{\sqrt{2}}Z k_{T,P}\gamma_0\gamma_5\rho_N(\textbf{r}),
 \label{Htp}
\end{eqnarray}
where $G_F$ is the Fermi constant, $\gamma_0$ and $\gamma_5$ are the Dirac matrixes and $\rho_N(\textbf{r})$ is the nuclear density normalized to unity.
To extract the fundamental $k_{T,P}$ constant from an experiment one needs to know an electronic structure factor, $W_{T,P}$, on a nucleus of interest:
%. $W_{T,P}$ is a characteristic factor for a given atom in a molecule
\begin{equation}
\label{WTP}
W_{T,P} = \frac{1}{\Omega}
\langle \Psi|\sum_i\frac{H_{T,P}(i)}{k_{T,P}}|\Psi\rangle.
\end{equation}
Similarly to \Eeff, the $W_{T,P}$ parameter cannot be measured and have to be obtained from a molecular electronic structure calculation.

In Refs.~\cite{Skripnikov:14a,FDK14} is was demonstrated that $^{229}$ThO molecule can be used to search for T,P-odd interaction of $^{229}$Th nuclear magnetic quadrupole moment (MQM) with electrons.
The T,P-odd electromagnetic interaction is described by the Hamiltonian \cite{Kozlov:87,Ginges:04}
\footnote{The T,P-violating magnetic quadrupole moment (\ref{eqaux1}) gives rise to a vector potential, $\vec{A}^{MQM}$, see Eqs. (165-167) in Ref.~\cite{Ginges:04}. Substituting $\vec{A}^{MQM}$ to the Dirac equation we go to the interaction $|e| (\vec{\alpha} \cdot \vec{A}^{MQM})$ coinciding with Eq.~(\ref{hamq}).}:
 \begin{align}\label{hamq}
% H  &=
 H^{MQM}  &=
 -\frac{  M}{2I(2I-1)}  T_{ik}\frac{3}{2} \frac{[\bm{\alpha}\times\bm{r}]_i r_k}{r^5},
 \end{align}
where Einstein's summation convention is implied, $\bm{\alpha}$ are the 4x4 Dirac matrices,
$ \bm{\alpha}=
  \left(\begin{array}{cc}
  0 & \bm{\sigma} \\
  \bm{\sigma} & 0 \\
  \end{array}\right),
$
$\bm{r}$ is the displacement of the electron from the Th nucleus, $\bm I$ is the nuclear spin,  $M$ is the nuclear MQM,
\begin{align}\label{eqaux1}
M_{i,k}=\frac{3M}{2I(2I-1)}T_{i,k}\, \\
 T_{i,k}=I_i I_k + I_k I_i -\tfrac23 \delta_{i,k} I(I+1)\,.
 \end{align} 
In the subspace of $\pm \Omega$ states Hamiltonian (\ref{hamq}) is reduced to the following effective molecular Hamiltonian~\cite{Sushkov:84}:
 \begin{align}\label{eq0}
% H_\mathrm{eff} &=
H^{MQM}_\mathrm{eff} &=
 -\frac{W_M  M}{2I(2I-1)} \bm S \hat{\bm T} \bm n
 \,,
 \end{align}
where $\bm S$ is the effective electron spin~\cite{KL95}, $S{=}|\Omega|{=}1$.
$W_M$ parameter can be evaluated by the following matrix element~\cite{FDK14}:
\begin{align}
  \label{WM}
W_M= 
\frac{3}{2\Omega} 
   \langle
   \Psi\vert\sum_i\left(\frac{\bm{\alpha}_i\times
\bm{r}_i}{r_i^5}\right)
 _\zeta r_\zeta \vert
   \Psi\rangle\ .
 \end{align}
It was shown in Refs.~\cite{Skripnikov:14a,FDK14} that using the $^{229}$Th isotope one can obtain limits on the strength constants of T,P-odd nuclear forces, neutron EDM, QCD vacuum angle $\theta$, quark EDM and chromo-EDM. This also can be applied to $^{229}$ThF$^+$ cation.

A commonly used way of verification the theoretical \Eeff, $W_{T,P}$ and $W_M$ values is to calculate ``on equal footing'' (using the same approximation for the wave function) those molecular characteristics (properties or effective Hamiltonian parameters) which have comparable sensitivity to different variations of wave function but, in contrast, can be measured. Similar to \Eeff, $W_{T,P}$ and $W_M$ these parameters should be sensitive to a change in the spin-polarized share of the electronic density, etc., in the atomic core region. The hyperfine structure (HFS) constant, $A_{||}$, is traditionally used as such a parameter (e.g., see Ref.~\cite{Kozlov:97}). To obtain $A_{||}$ on $^{229}$Th in the $^{229}$ThF$^+$ theoretically, one can evaluate the following matrix element:
\begin{equation}
 \label{Apar}
A_{||}=\frac{\mu_{\rm Th}}{I\Omega}
   \langle
   \Psi|\sum_i\left(\frac{\bm{\alpha}_i\times
\bm{r}_i}{r_i^3}\right)
_\zeta|\Psi
   \rangle, \\
\end{equation}
where $\mu_{\rm Th}$ is magnetic moment of an isotope of $^{229}$Th nucleus having spin $I$. In the present paper we do not consider fluorine nuclear spin.

For the preparation and conduction of the experiment the value of g-factor of the molecule is of interest. It is defined as 
\begin{eqnarray}
 \label{Gpar}
G_{\parallel} &=&\frac{1}{\Omega} \langle \Psi |\hat{L}^e_{\hat{n}} - g_{S} \hat{S}^e_{\hat{n}} |\Psi \rangle,  
\end{eqnarray}
where ${\vec{L}}^e$ and ${\vec{S}}^e$ are the electronic orbital and electronic spin momenta operators, respectively; $g_{S} = -2.0023$ is a free$-$electron $g$-factor. Note that the value of $G_{\parallel}$ is close to zero for the $^3\Delta_1$ state (and equal to zero when both the scalar-relativistic approximation is applied and the radiation corrections to the free-electron g-factor are ignored. Therefore, the parameter is very sensitive to the quality of the wave function, since high-order interference contributions between the spin-orbit and electron correlation effects become important.

Recently, ThF$^+$ has been studied both experimentally and theoretically in Refs.~\cite{Barker:2012, Heaven:14}. The measured and calculated values are given there for spectroscopic constants of the lowest-lying states including $^3\Delta_1$. The latter is found to be the first excited state. However, up to now there is only one (semiempirical) estimate of \Eeff\ in ThF$^+$ published in Ref.~\cite{Meyer:08}, \Eeff=90 GV/cm. The aim of the paper is to perform accurate ab-initio study of ThF$^+$ electronic structure and calculate \Eeff\  and other parameters given by Eqs.~(\ref{WTP},\ref{WM},\ref{Apar}).

%############################
\section{Two-step approach}
\label{s2step}
%############################

%def:    In general, the core characteristics are measurable properties and other effective Hamiltonian parameters which are sensitive to variation of electronic wave function (electronic densities) in atomic core regions though the core variation itself is induced by changes (perturbation, excitation) of the wave function in the valence region (by environment, chem. bonding, external fields, etc.)

It follows from Eqs.~(\ref{Wd})--(\ref{Apar}) that the action of operators related to the \Eeff, $W_{T,P}$, $W_M$ and $A_{||}$ characteristics is heavily concentrated in the atomic core region. On the other hand the leading contribution to the corresponding matrix elements (mean values) is due to the valence electrons since contributions from the inert (usually closed and spherically symmetric) inner-core shells compensate each other or negligible in most cases of practical interest for the operators, in particular, dependent on the total angular momentum and spin. Note, however, that the spin-polarization of core (sub-valence or outer-core) shells induced by the valence unpaired electrons can provide a comparable contribution by magnitude to such properties as that from the valence electrons, e.g., see Refs.~\cite{Mosyagin:98, Kudashov:13, Skripnikov:11a}. Below we shall call such properties as the ``core properties'' (or, more generally, ``core characteristics'' since not only measurable properties but other effective Hamiltonian parameters which are not always measurable can be considered here) assuming that the main contribution to them comes from the spatially-localized core region rather than from core shells. Some of well-known examples of such property are the magnetic dipole hyperfine constants (see Eq.~\ref{Apar}). In the cases of unpaired $s$-electrons, the leading contribution to the hyperfine structure is determined by the Fermi-contact interaction
(in nonrelativistic case)
which is proportional to the electronic spin density directly on the nucleus. Other examples of core properties (which have negligible contribution from inert inner-most core shells) are the chemical shifts of X-ray emission spectra \cite{Lomachuk:13, Titov:14a}, etc.

One can safely exclude inactive inner-core electrons from correlation calculation due to their negligible contribution to the core properties. In the present consideration, the inner-core consists of $1s{-}4f$ electrons of Th. As Th is a very heavy element (atomic number is 90) the interaction of electrons with the Th nucleus should be treated by a fully relativistic manner for a good accuracy. Moreover, for some properties even taking account of Breit interaction (mainly between valence and core electrons of Th, see \cite{Petrov:04b, Mosyagin:06amin}) can be important. With a good accuracy for the properties considered here, the inner-core electrons differ negligibly in the cases of atomic Th and ThF$^+$ cation because their wavefunctions mostly defined by the strong Th nucleus potential $\sim \frac{Z}{r}$  screened by inner-more electrons, so that the effective Th core field is much stronger than the energetics of valence (chemically active) electrons. In the correlation calculation they can be frozen without a loss of the accuracy accessible presently. A common way to exclude inner-core electrons is to use the relativistic effective core potential method. Earlier our group has developed the generalized  relativistic effective core potential (GRECP) version which permits one to attain a very high accuracy \cite{Titov:91, Titov:99, Mosyagin:10a}. This effective potential emulates interaction between inner-core electrons (excluded explicitly from GRECP calculations) and valence plus outer-core electrons (treated explicitly with GRECP).

Performing electronic structure calculation one can evaluate different valence properties such as transition energies between low-lying states, molecule-frame dipole moments, etc. However, since the inner-core parts of the valence one-electron ``pseudo-wavefunctions''
% \footnote{Molecular spinors in the (fully) relativistic case or orbitals in the scalar-relativistic and nonrelativistic cases.}
%
are smoothed in the GRECP calculations, they have to be recovered with some core-restoration method before using them to evaluate the core characteristics considered above. In series of papers a non-variational restoration concept (and its initial implementation, see \cite{Titov:06amin} and references), which is based on a proportionality of valence and low-lying virtual spinors in the inner-core region of heavy atoms was developed (see \cite{Titov:14a} and the next section).
Recently we have developed a new implementation of the concept which permits to use well-developed codes on correlation treatment such as {\sc dirac} \cite{DIRAC12}, {\sc mrcc} \cite{MRCC2013} and  {\sc cfour} \cite{CFOUR} \cite{Skripnikov:11a, Skripnikov:13b}.
Below we give description of the new implementation; now the code is also extended to characterize effective states (configurations) of atoms in compounds, e.g. Th in the ThF$^+$ and ThO.

Using the basic idea of the nonvariational restoration method one generates {\it equivalent} basis sets of one-center four-component spinors 
$$
  \left( \begin{array}{c} f_{nlj}(r)\theta_{ljm} \\
     g_{nlj}(r)\theta_{2j{-}l,jm} \\ \end{array} \right)
$$
and smoothed two-component pseudospinors 
$$
    \tilde f_{nlj}(r)\theta_{ljm}
$$
in all-electron finite-difference Dirac-Fock-Breit and GRECP\,/\,self-consistent field calculations (employing the $jj-$coupling scheme) of {\it the same} configurations of a considered atom and its ions \cite{HFDB, Bratzev:77, HFJ, Tupitsyn:95}
\footnote{These sets, describing mainly the given atomic core region, are generated independently of the basis set used for the molecular (GRECP) calculations.}.
Here $n$ is the principal quantum number, $j$ is the total electronic momentum, $m$ is its projection and $l$ is the orbital momentum. In the newly developed procedure a basis set of real spin-orbitals (and not complex spin-orbit-mixed spinors) $\tilde{\xi}_p$ is generated additionally.
The spin-orbitals $\tilde{\xi}_p$ are then expanded in the basis set of one-center two-component atomic {pseudospinors}
\begin{equation}
   \tilde{\xi}_p \approx
    \sum_{l=0}^{L_{max}}\sum_{j=|l-1/2|}^{j=|l+1/2|} \sum_{n,m}
    T_{nljm}^{p}\tilde f_{nlj}(r)\theta_{ljm}\ .
 \label{expansion}
\end{equation}

The atomic two-component pseudospinors are replaced by equivalent four-component spinors while the expansion coefficients from Eq.~(\ref{expansion}) are preserved:
\begin{equation} 
%  \tilde{\xi}_p(\mathbf{x}) 
  \xi_p
  =
    \sum_{l=0}^{L_{\rm max}}\sum_{j=|l-1/2|}^{j=|l+1/2|} \sum_{n,m}
    T_{nljm}^{p}
     \left(
    \begin{array}{c}
    f_{nlj}(r)\theta_{ljm}\\
    g_{nlj}(r)\theta_{2j-l,jm}
    \end{array}
    \right)
 \label{restoration}
\end{equation}
and we obtain four-component function $\xi_p$ which is ``equivalent'' to $\tilde{\xi}_p$.
If a one-electron reduced density matrix with elements  $\tilde{P}_{\mu\nu}$ in a basis set of multi-center spinors (or spin-orbitals) $\psi_\mu$  is evaluated after the (G)RECP calculation of a molecule or some condensed-matter system (see Ref.~\cite{Skripnikov:13b} for details on the condensed-matter case) one can then reexpand it in the basis of one-center $\tilde{\xi}_p$ functions on an atom of interest. This mapping from a multi-center basis $\{\psi_\mu\}$ to the one-center basis $\{\tilde{\xi}_p\}$ corresponds to a similarity transformation of the density matrix:
\begin{equation} 
%   \tilde{P}_{\mu\nu} \longrightarrow \tilde{D}_{pq}
||\tilde{P}_{\mu\nu}|| \longrightarrow ||\tilde{D}_{pq}||
 \label{transf1}
\end{equation}
where $\tilde{D}_{pq}$ are elements of the density matrix in the basis of $\tilde{\xi}_p$ functions. Due to ``equivalence'' of $\tilde{\xi}_p$ and $\xi_p$ functions (see Eqs.~(\ref{expansion}) and (\ref{restoration})) based on appropriate properties of the hard-core shape-consistent (G)RECP versions \cite{Titov:02Dis}, one can write:
\begin{equation} 
%  D_{pq} \approx \tilde{D}_{pq},
||D_{pq}|| \approx ||\tilde{D}_{pq}||
 \label{transf2}
\end{equation}
where $D_{pq}$ are elements of the density matrix in the basis of four-component $\xi_p$ functions~(\ref{restoration}). Thus, as an approximation we can equate the $D_{pq}$ elements to $\tilde{D}_{pq}$ and interpret it as a restoration of ``true'' four-component structure of density matrix that is important first of all for the inner-core region.

The mean value of some one-electron operator $A$ corresponding to a core property on a given atom can be evaluated as follows:
\begin{equation} 
   \langle {A} \rangle\ =\  \sum_{pq} D_{pq}    {A}_{pq}\ ,
 \label{CoreProp} 
\end{equation} 
where ${A}_{pq}$ are the matrix elements of operator $A$ in the basis of four-component functions $\xi_p$~(\ref{restoration}).  

In the current implementation of restoration procedure the functions $\tilde{\xi}_p$ are real spin-orbitals with the spatial factor in the form of contracted Gaussians, for which reexpansion (\ref{transf1}) is performed analytically. Therefore, significant acceleration is  attained in contrast to the original restoration procedure \cite{Titov:06amin}. Note, however, that the four-component functions used to evaluate matrix elements of operator $A$ are taken in numerical (finite-difference) form. This permits to exclude some complications in reproducing accurate wavefunction behavior in a region near nucleus which can arise when Gauss-type functions are used there.

%######################################
\section{``Atoms in compounds'' theory}
%######################################

In Ref.~\cite{Titov:14a} we have introduced a concept of {\it atoms in compounds} (AIC) and applied it to the problem of chemical shifts of X-ray emission lines. The concept assumes that for the core characteristics one can determine an effective state of a given atom in a chemical compound, for which the mean values of operators corresponding to \textit{all} the considered core characteristics, have near the same magnitudes for the case of an atom bonded in a molecule and for the same atom in the considered effective state. Note, that the concept cannot be directly applied to evaluation of the ``valence'' properties (again, taking in mind spatial localization rather than affiliation to valence shells) such as the molecule-frame dipole moment, g-factor, etc.
 
Let us show how the AIC theory can be formulated in context of the problems discussed in this paper.  Assume that we have obtained a one-electron density matrix from calculation of an atom, molecule or crystal (in the direct lattice). The density matrix can be formally reexpanded on one center, i.e., on a heavy atom of interest:
\begin{equation}
\begin{array}{l}                 
\rho(\vec{r}|\vec{r'})= \\ \sum\limits_{nljm,n'l'j'm'} \rho_{nljm,n'l'j'm'} \varphi_{nljm}(\vec{r})\varphi_{n'l'j'm'}^{\dagger}(\vec{r'})
 \label{rrofull}
\end{array}
\end{equation}
in a sufficiently complete basis set of orthonormal atomic functions $\{\varphi_{nljm}\}$. Then for the mean value of some one-electron operator $A$ we have:
\begin{equation} 
   \langle {A} \rangle\ =\  \sum_{nljm,n'l'j'm'} \rho_{nljm, n'l'j'm'}\int \varphi_{n'l'j'm'}^{\dagger}A\varphi_{nljm}d\vec{r}\ .
 \label{OperFull} 
\end{equation} 
The mean value in Eq.~(\ref{OperFull}) can be rewritten as
\begin{equation}
\begin{array}{r}                 
  \int\varphi_{n'l'j'm'}^{\dagger}A\varphi_{nljm}d\vec{r}=\int\limits_{|\vec{r}|\le|\vec{R_c}|} \varphi_{n'l'j'm'}^{\dagger}A\varphi_{nljm}d\vec{r} \\
  +\int\limits_{|\vec{r}|>|\vec{R_c}|} \varphi_{n'l'j'm'}^{\dagger}A\varphi_{nljm}d\vec{r}\ ,
 \label{MatrFull} 
\end{array}
\end{equation}
where $R_c$ is some ``core radius'' (see also below). Here we consider that the operator $A$ corresponds to a core property. It means that for $r{>}R_c$ the second term in Eq.~(\ref{MatrFull}) have to be negligible compared to the first term.  As an extremal case, $R_c{=}0$ for the Fermi-contact interactions. Thus, for a core-property operator we have 
\begin{equation}
\begin{array}{r}                 
  \int\varphi_{n'l'j'm'}^{\dagger}A\varphi_{nljm}d\vec{r} \approx \int\limits_{|\vec{r}|\le|\vec{R_c}|} \varphi_{n'l'j'm'}^{\dagger}A\varphi_{nljm}d\vec{r}\ .
 \label{MatrRed} 
\end{array}
\end{equation}

Now we assume that the basis set $\{\varphi_{nljm}\}$ was constructed after calculations of the atom or its low-charged ions.  The basis set contains inner core spinors, marked below by index ``C'' (which are occupied by the inert electrons and excluded from molecular calculations with GRECP as inactive, completely occupied states); valence, outer core (occupied by explicitly treated core electrons) and low-lying virtual spinors; all together they are marked by index ``W''. The rest spinors, corresponding to high-energy virtual states, are marked by index ``R''.
The core states are only negligibly changed in the low-energy process under consideration (formation of chemical bond, low-energy excitation of atom, etc.). 
The completeness condition for the $\{\varphi_{nljm}\}$ basis can be formally written as
\begin{eqnarray}
1=\sum\limits_{nljm} \varphi_{nljm}\varphi_{nljm}^{\dagger}=P_C+P_W+P_R, \ \\
P_C=\sum\limits_{nljm\in C} \varphi_{nljm}\varphi_{nljm}^{\dagger}, \ \\
P_W=\sum\limits_{nljm\in W} \varphi_{nljm}\varphi_{nljm}^{\dagger}, \ \\
P_R=\sum\limits_{nljm\in R} \varphi_{nljm}\varphi_{nljm}^{\dagger}, \
%end ls5
\end{eqnarray}
where $P_C$ is the projector on the inner-core spinors, $P_W$ is the projector on the outer-core, valence and low-energy virtual spinors, and $P_R$ is the projection on the other (high-energy)   states. For low-energy processes, which include chemical bonding, low-lying excitations and those induced by weak external fields, one can usually neglect the high-energy states to study the properties of interest:
\begin{equation}
\begin{array}{r}   
   \rho=(P_C+P_W+P_R)\rho(P_C+P_W+P_R) \\ 
   \approx (P_C+P_W)\rho(P_C+P_W) \\
   \approx P_C \rho P_C + P_W \rho P_W=\rho^C+\rho^W\ ,
 \label{RhoRed} 
\end{array} 
\end{equation} 
where $\rho^C= P_C \rho P_C$ and $\rho^W= P_W \rho P_W$. Here we have taken into account that the inner-core electrons need not be usually correlated to preserve high accuracy for the core properties in general, in contrast to the W-states. Therefore, the off-diagonal blocks $P_W\rho P_C$ and $P_C\rho P_W$ can be mostly  neglected
   \footnote{Note, however, that relaxation of the inner-core shells can notably influence on some core characteristics in particular cases and should be taken into account using the variational core restoration procedures \cite{Titov:96}; such cases will be considered elsewhere.}.

Due to Eq.~(\ref{RhoRed}) the expression (\ref{OperFull}) for $\langle{A}\rangle$ reduces to the following two terms:
\begin{equation}
\begin{array}{r}  
\langle {A} \rangle\ \approx \langle {A} \rangle^C + \langle {A} \rangle^W,
 \label{OperRed} 
\end{array}
\end{equation}
where
\begin{equation}
\begin{array}{l}  
   \langle {A} \rangle^C = \\
   \sum\limits_{nljm\in C,n'l'j'm'\in C} \rho_{nljm, n'l'j'm'}\int \varphi_{n'l'j'm'}^{\dagger}A\varphi_{nljm}d\vec{r}\ ,
\\ 
\\
\langle {A} \rangle^W = 
\\
   \sum\limits_{nljm\in W,n'l'j'm'\in W} \rho_{nljm, n'l'j'm'}\int \varphi_{n'l'j'm'}^{\dagger}A\varphi_{nljm}d\vec{r}\ .\\
\end{array}
\end{equation}

For such properties as hyperfine structure constant, etc.\ considered here $\langle {A} \rangle^C \approx 0$, i.e., direct contribution from the closed-shell core electrons can be ignored (and only their spin-polarization by open valence shells can be not negligible, see \cite{Skripnikov:11a}). In other cases $\langle {A} \rangle^C$ can be obtained from atomic calculation if one takes into account that the heavy-atom inner-core electrons are inactive in low-energy processes.
Thus, for our case we have:
\begin{equation} 
\begin{array}{r} 
   \langle {A} \rangle \approx \langle {A} \rangle^W
 \label{OperRedFin}\ . 
\end{array} 
\end{equation} 

\begin{figure}[]
\includegraphics[width = 3.3 in]{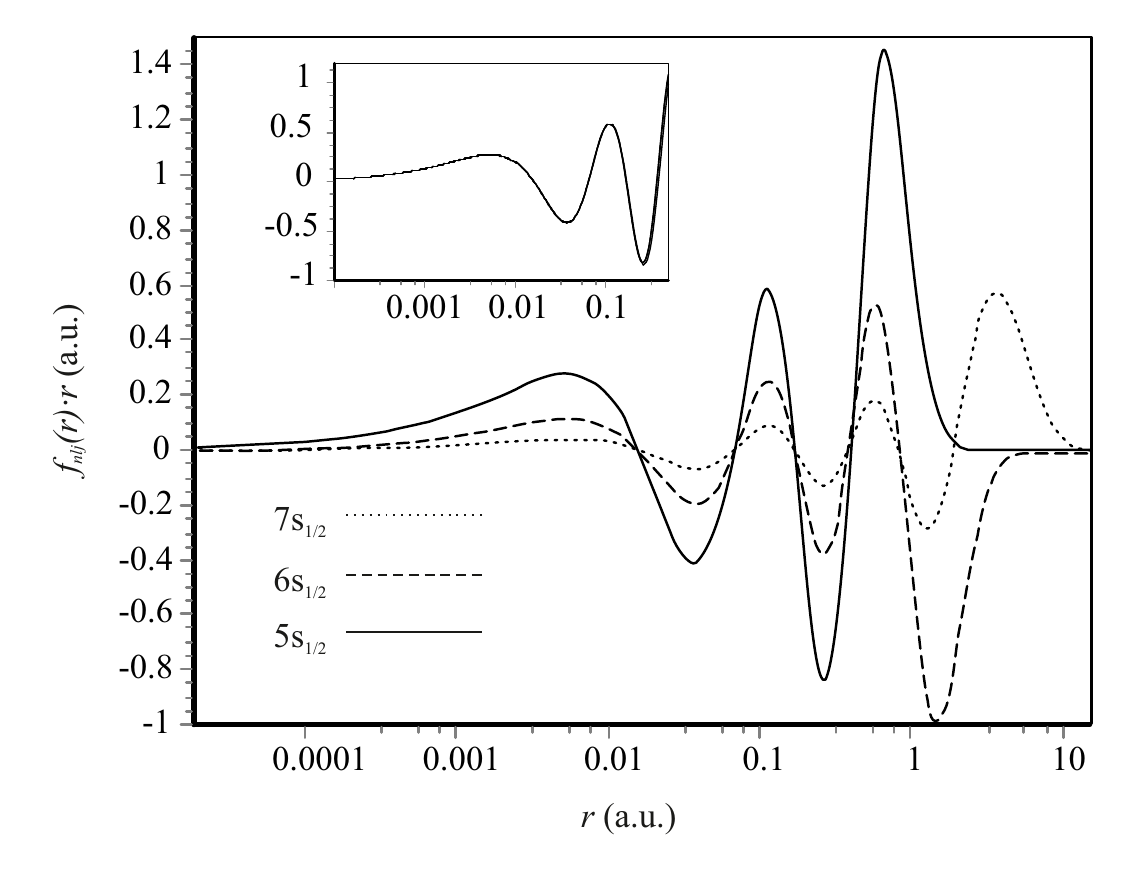}
 \caption{Large components of the $5s_{1/2}$, $6s_{1/2}$ and $7s_{1/2}$  spinors of Th for the $7s^27p^16d^1$ configuration.  The large components of $5s_{1/2}$, $6s_{1/2}$ and $7s_{1/2}$ spinors in the core region are given in subfigure, where the scaling factor is chosen in such a way that the amplitudes of large components of these spinors are equal at $R_c$ = 0.25~a.u.}
 \label{Th}
\end{figure}

It is well known for heavy atoms and their compounds \cite{Petrashen:56, Flambaum:90, Titov:99, Titov:02Dis} that the valence one-electron wave-functions and low-lying virtual states are proportional to each other in the vicinity of a nucleus. This is due to overwhelming contribution of highly charged (even being shielded) heavy-nucleus potential as compared to other potentials from molecular environment, inter-electron interaction, etc. In figure \ref{Th} one can see the large components of $5s_{1/2}$, $6s_{1/2}$ and $7s_{1/2}$ of Th atom taken from a self-consistent field calculation of the $7s^27p^16d^1$ configuration. Note that core radius $R_c$ belongs to the region of proportionality for the considered here core properties.

%=============
%===========

One can introduce some reference functions for each combination of $l,j$:
\begin{eqnarray}
  {\cal{H}}_{ljm}(\vec{r}) \equiv
  \left( \begin{array}{c} \eta^f_{lj}(r)\theta_{ljm} \\
     \eta^g_{lj}(r)\theta_{2j-l,jm} \\ \end{array} \right)\ .
 \label{eta}
\end{eqnarray}
   The functions $\eta^{f,g}_{lj}(r)$  will be determined such that they are equal to valence functions $f_{lj}(r), g_{lj}(r)$ of a given atomic four-component spinor with the same $lj$ for $r{<}R_c$ and equal to zero outside the (given core) region.  Due to the proportionality property (see figure~\ref{Th}) it is not practically important which of the  W-functions, $\varphi_{nljm}^W(\vec{r})$, is chosen for a given $lj$ or from which configuration it is chosen
   \footnote{Note in this connection that only one of two function, $\eta^f_{lj}(r)$ and $\eta^g_{lj}(r)$ can be considered as ``independent'', and choosing one of them automatically fixes the other one.
%new
We use the convention that phase of the $\eta^f_{lj}(r)$ function is chosen in such a way that the function is positive in its first maximum.
%end new   
   }.
For example, for the case of $l=0$, $j=1/2$ of Th one can consider $5s_{1/2}$, or $6s_{1/2}$, or $7s_{1/2}$ as follows from Fig.~\ref{Th}. For all functions from the W-diversity (i.e.\ for all $n{\in}W$)
\begin{eqnarray}
    \varphi_{nljm}(\vec{r}) \approx k_{nljm} {\cal{H}}_{ljm}(\vec{r})\ ,\ \  r\le R_c\ ,
 \label{proport}     
\end{eqnarray}
   where $k_{nljm}$ are the proportionality (scaling) factors.

Using Eqs. (\ref{MatrRed}) and (\ref{proport}) we can rewrite Eq.~(\ref{OperRedFin}) in the following form:
\begin{equation} 
\begin{array}{l} 
\langle {A} \rangle\ 
     {\approx}\sum\limits_{nljm; n'l'j'm'\in W} \rho_{nljm, n'l'j'm'}\int \varphi_{n'l'j'm'}^{\dagger}A\varphi_{nljm}d\vec{r}\\
     {\approx}\sum\limits_{nljm; n'l'j'm'\in W} \rho_{nljm, n'l'j'm'}k_{nljm} k_{n'l'j'm'}\int {\cal{H}}_{l'j'm'}^{\dagger}A{\cal{H}}_{ljm} d\vec{r} \\          
     = \sum\limits_{ljm;l'j'm'} \Delta_{ljm, l'j'm'} \int {\cal{H}}_{l'j'm'}^{\dagger}A{\cal{H}}_{ljm} d\vec{r}\ ,              
 \label{OperRedFinFin} 
\end{array}  
\end{equation} 
where
\begin{equation} 
\begin{array}{l} 
\Delta_{ljm, l'j'm'}=\sum\limits_{nljm; n'l'j'm'\in W} \rho_{nljm, n'l'j'm'}k_{nljm} k_{n'l'j'm'}\ .
 \label{WRed} 
\end{array}  
\end{equation} 
Here $||\Delta_{ljm, l'j'm'}||$ is the W-reduced density matrix, in which the terms are summed up on the chosen principal quantum numbers $n\in W$ in contrast to a conventional one-electron density matrix
from Eq.~(\ref{rrofull}).

The last expression in Eq.~(\ref{OperRedFinFin}) means that for calculation of a core property $A$ it is sufficient to know some W-reduced density matrix $\Delta_{ljm, l'j'm'}$ as well as matrix elements of the operator over the reference functions  $\{{\cal{H}}_{ljm}(\vec{r})\}$.
One can interpret $||\Delta_{ljm, l'j'm'}||$ as a density matrix of an effective AIC state. The diagonal elements of the matrix are occupancies of the reference functions
   \footnote{Note that if $\{\eta^f_{lj}(r)\}$ coincide by amplitude with some chosen atomic radial spinors $\{f_{n_0lj}(r)\}$ within $r{<}R_c$ (the same will be automatically for $\{\eta^g_{lj}(r)\}$ vs.\ $\{g_{n_0lj}(r)\}$), the diagonal elements can be interpreted as occupancies in terms of these spinors. Otherwise, if the four-component $\{{\cal{H}}_{ljm}(r)\}$ functions are normalized, as is used in \cite{Titov:14a}, the diagonal elements can be rather interpreted as the ``partial wave charges''.}.
The nondiagonal elements between different $lj$ are ``overlap occupancies (populations)'' The latter can occur in consideration of a molecule/crystal due to polarization of atomic orbitals in molecular/crystalline environment or in an atom placed in some external field, i.e.\ they reflect non-spherical distribution of electron density in the vicinity of nucleus of atom under consideration.
%   (or, in principle, in some ``exotic'' state of atom, e.g.\ in the case of accidental degeneracy).
We should stress that the diagonal and overlap populations have meaning of ``observable quantities'' (though, in practice, their combination can be rather observed experimentally). It means that different parts of the W-reduced density matrix can be obtained from different experiments (or their combinations). In particular, the diagonal matrix elements can be    extracted from the X-ray emission chemical-shift experiments. The non-diagonal matrix elements  
 predetermine the value of effective electric field which, in turn, can formally be extracted from the electron EDM experiments if we know the \eEDM\ value.
Finally, one can say that the AIC effective state in some sense is a more general term than the classical effective state term since the AIC effective state can include overlap populations between different harmonics  as is discussed above.
%It means that we take into account non-spherical distribution of electron density in the core region of atom in compound.

The AIC concept described here can be applied both in the direct four-component calculation and in the two-step study, in which the four-component density matrix is obtained at the second stage of the procedure discussed in the previous section.
Actually, computation of the effective W-reduced density matrix is a special case of the recovery procedure when the equivalent basis sets are constructed only from the reference functions $\{{\cal{H}}_{ljm}(\vec{r})\}$ and some modification of one-center restoration is applied. In the present paper we report implementation of the procedure and its application to calculation of the W-reduced density matrix for the ThO and ThF$^+$ molecules and a number of core-properties: hyperfine magnetic dipole constant~(\ref{Apar}), effective electric field~(\ref{Wd}), the molecular-structure parameters of T,P-odd pseudoscalar-scalar electron-nucleus interaction (\ref{WTP}) and T,P-odd interaction of the nuclear magnetic quadrupole moment with electrons~(\ref{WM}). The code is interfaced to the {\sc dirac12} \cite{DIRAC12} and {\sc mrcc} \cite{MRCC2013} codes.

%##############################
\section{Computational details}
%##############################

To evaluate \Eeff, $W_{T,P}$, $W_M$ and $A_{||}$ in ThF$^+$ we have applied the two-step method described above. The computational scheme used in the present paper is similar to that employed in \cite{Skripnikov:13c, Skripnikov:14b} for calculation of ThO, where we have described and analysed the scheme in details (possible sources of errors, importance of correlation treatment, importance of multireference approaches, applicability and convergence of multireference configuration interaction approaches, etc.).  In all the calculations the $1s{-}4f$ inner-core electrons of Th were excluded from molecular correlation calculations using the valence (semi-local) version of the   GRECP \cite{Mosyagin:10a} method. The main calculation was performed within the 38-electron two-component single-reference coupled-cluster method with single, double and perturbative triple cluster amplitudes, 38e-2c-CCSD(T). The calculation was perform using MBas basis set, generated in \cite{Skripnikov:14b} with added $h$ and $i$ type functions, i.e., we used the (30,20,10,11,4,6,5)/[30,8,10,4,4,2,1] basis set. For F we have applied the aug-ccpVQZ basis set \cite{Kendall:92} with two removed $g$-type basis functions, i.e., the (13,7,4,3)/[6,5,4,3] basis set was used. To consider high-order correlation effects we calculated correlation correction. For this we have frozen 20 outer core electrons ($5s^2 5p^6 5d^{10}$ shells of Th and $1s^2$ shell of F) and performed two-component calculations within the coupled-cluster method with single, double, triple and perturbative quadruple cluster amplitudes, CCSDT(Q), and within the CCSD(T) method. We utilized
the CBasSO atomic natural basis set which was generated using the same procedure that was used and described in \cite{Skripnikov:14b, Skripnikov:13a} and can be written as (35,29,15,10,7)/[6,8,5,3,2] for Th, and (13,7)/[4,3] for fluorine. The correction was calculated as a difference between the calculated parameters within the CCSDT(Q) and CCSD(T) methods. In addition, the basis set enlargement corrections to the considered parameters were also calculated. For this we have performed: (i) scalar-relativistic CCSD(T) calculation using the same basis set as used for the main two-component calculation; (ii) scalar-relativistic CCSD(T) calculation utilizing the extended basis set on Th (Lbas basis set (37,29,15,14,10,10,5)/[22,17,15,14,10,10,5] generated in \cite{Skripnikov:14b}). Corrections were estimated as differences between the values of the corresponding parameters.
Finally, we have calculated the vibrational contribution to the considered core properties and molecule-frame dipole moment corresponding to zero vibrational level of the $^3\Delta_1$ electronic state as a difference between the value averaged over zero vibration wave function and the non-averaged value at the given internuclear distance (3.75 a.u., see below). The potential energy curve was calculated at the 38-electron one-component CCSD(T) level with the LBas basis set.

%################################
\section{Results and discussions}
%################################

According to the 38-electron two-component CCSD(T) calculations the equilibrium internuclear distance in the $^3\Delta_1$ state of ThF$^+$ is 3.75 a.u.\ which agrees well with the experimental datum \cite{Barker:2012}, see table \ref{spec_props}. In calculations of the parameters under consideration we have set R(Th--F) to 3.75 a.u.

\begin{table}[b]
  \caption{Equilibrium internuclear distance $R_e$, harmonic vibrational wavenumber $\omega_e$ and vibrational anharmonicity $\omega_e x_e$ for the $^3\Delta_1$ state of ThF$^+$.}
  \begin{tabular}{lccc}
   \hline \hline
   Method                & $R_e$, a.u. & $\omega_e$, cm$^{-1}$& $\omega_e x_e$, cm$^{-1}$ \\
   \hline   
   MRCI+Q/SO, \cite{Barker:2012}     & 3.76    & 655.6     & ----	     \\      
   CCSD(T), this work    & 3.75  & 658.4 & 1.9 \\
   Experiment, \cite{Barker:2012}    & 3.74(4) & 658.3(10) & ----	     \\   
   \hline \hline
  \end{tabular}
  \label{spec_props}
\end{table}

Table \ref{TResultsEeff} lists the calculated values of effective electric field along with the parameter of the T,P-odd pseudoscalar$-$scalar electron$-$nucleus neutral currents interaction, hyperfine structure constant, $W_M$ parameter and g-factor for the $^3\Delta_1$ state of ThF$^+$.
It follows from table \ref{TResultsEeff} that the calculated value of \Eeff\ is stable with respect to the electron correlation improvement and basis set enlargement. 
Similar to Ref.~\cite{Skripnikov:14b} using the size-extensive coupled-cluster calculations we have found that the outer-core electrons of Th contribute about 3.5 GV/cm to \Eeff\ (similar value was found in Ref.~\cite{Skripnikov:14b} for ThO) and -161 $\frac{\mu_{\rm Th}}{\mu_{\rm N}}\cdot$MHz to A$_{||}$. Thus, if one performs 18-electron rather than 38-electron calculation the outer-core contributions should be taken in mind.
According to our calculations the spin-orbit contribution from the valence electrons to \Eeff\ (about 1~GV/cm) is almost negligible in the case of ThF$^+$, in contrast to ThO where it is about 10 GV/cm \cite{Skripnikov:14b}). According to table \ref{TResultsEeff} in view of the extensive analysis of uncertainties performed in Ref.~\cite{Skripnikov:14b} we suggest that the theoretical uncertainties of \Eeff, $W_{T,P}$ and A$_{||}$ are within 7\%.
Unfortunately, A$_{||}$ (Eq.~\ref{Apar}) is yet unknown experimentally for $^{229}$ThF$^+$ and it cannot be used currently to check the value of \Eeff\ and other considered properties. However, we have shown in Ref.~\cite{Skripnikov:14c} for the ground state of PbF molecule that the used computational scheme is rather accurate: the calculated value of A$_{||}$(PbF) agrees with the experimental datum \cite{Mawhorter:11} within 2\%.
Finally, it should be noted that the estimation made in Ref.~\cite{Meyer:08} for \Eeff\ (90 GV/cm) is more than twice overestimated (similarly strong overestimation was also found for \Eeff\ in PtH$^+$, see \cite{Skripnikov:09}). In a like manner the estimations for the $W_{T,P}$ and $W_M$ parameters made in Refs.~\cite{Dzuba:11,FDK14} and based on \Eeff\ from Ref.~\cite{Meyer:08} are also about twice overestimated.

The effective electric field in the $^3\Delta_1$ state of ThF$^+$ is about two times smaller than the \Eeff\ in the $^3\Delta_1$ state of ThO (81.5~GV/cm, see \cite{Skripnikov:14b}) because of a smaller mixing of $s$ and $p$ orbitals. We can give the following explanation. 
%ls1
%The molecule-frame dipole moment of ThO is about 1.5 times bigger than the dipole moment of ThF$^+$ (with respect to the Th nucleus), see table \ref{TResultsEeff}. 
%Thus, the charge transfer from thorium to the light atom (O or F) is higher in ThO than in ThF$^+$. 
In the naive ionic model ThO can be considered as Th$^{+2}$ and O$^{-2}$,
ThF$^+$ can be considered as Th$^{+2}$ and F$^{-1}$.
This agrees with the fact that the molecule-frame dipole moment of ThO is about 1.5 times 
larger than the dipole moment of ThF$^+$ with respect to the Th nucleus (see Table II).
%end ls1
This leads to higher effective negative electric charge on oxygen in ThO than on fluorine in ThF$^+$. Both ThO and ThF$^+$ have two unpaired electrons. They are non-bonding and localized on Th so that Th has $\sigma^1\delta^1$ configuration in both cases, where $\sigma$ is mainly the $7s$ atomic orbital of Th and $\delta$ in mainly $6d$ atomic orbital of Th. The unpaired electrons of Th feel a stronger electric field in ThO than in ThF$^+$. This leads to higher polarization of the unpaired electrons in the case of ThO, i.e., stronger $s{-}p$ mixing of $\sigma$-state ($\delta$ state has no practical interest for \Eeff\ here due to far smaller amplitude of $6d$ in the core region than $7s$). The leading contribution to
\Eeff\ is roughly proportional to 
 $$C_{7s}C_{7p}\langle 7s|H_d/d_e|7p\rangle\ ,$$
where Eq.~(\ref{matrelem}) is used. The matrix element is mainly accumulated near the Th nucleus and $C_{7s}({\approx1}), C_{7p}$ are the corresponding MO~LCAO coefficients of the atomic Th orbitals in the hybridized molecular one. As a consequence the effective electric field should be expected notably larger in ThO vs.\ ThF$^+$. On the other hand the smaller polarization of open-shell $\sigma$-state leads to higher $s-$character of the orbital and the hyperfine structure constant in $^3\Delta_1$ of ThF$^+$ is bigger than in ThO (see table \ref{TResultsEeff}). Note, that the hyperfine structure constant behaves ``inconsistently'' with respect to the effective electric field in the present case.

\begin{table*}[!h]
\caption{
The calculated values of the molecule-frame dipole moment ($d$), effective electric field (\Eeff), parameter of the T,P-odd pseudoscalar$-$scalar electron$-$nucleus neutral currents interaction ($W_{T,P}$), parameter of T,P-odd MQM interaction ($W_M$), hyperfine structure constant (A$_{||}$) and g-factor ($G_{\parallel}$) of the $^3\Delta_1$ state of ThF$^+$ compared to the corresponding values of ThO from Ref.~\cite{Skripnikov:14a,Skripnikov:14b} using the coupled-cluster methods. 
}
\label{TResultsEeff}
\begin{tabular}{l l  r  r  r  r  r  r  c}
\hline\hline
 Method   & $d$ $^*$,  & \Eeff, & $W_{T,P}$, & $W_M$ & A$_{||}$,   & $G_{\parallel}$  \\
            & Debye & GV/cm & kHz & $\frac{10^{33}\mathrm{Hz}}{e~{\rm cm}^2}$ & $\frac{\mu_{\rm Th}}{\mu_{\rm N}}\cdot$MHz & \\

\hline
  38e-2c-CCSD                 & {}  2.69   & {} 35.5  & 48 & 0.87 &  -4214 & 0.039~ \\  
  38e-2c-CCSD(T)              & {}  2.66   & {} 38.1  & 51 & 0.90 &  -4164 & 0.033~ \\ 
  correlation correction      & {}  0.07   & {} 0.0   & 0  &-0.01 &   13   & 0.001  \\    
  basis set correction        & {} -0.01   & {}-0.6   &-1  &-0.02 &  -14   & ---    \\ 
  vibr. contribution          & {}  0.03   & {}-0.1   & 0  &       &   2    & ---   \\
  \textbf{FINAL(ThF$^+$)}     & {}  2.74   & {} 37.3  & 50 & 0.88 &  -4163 & 0.034~ \\ 

\\
\hline
  \textbf{FINAL(ThO)}         & {}  4.23   & {} 81.5  & 112& 1.66   & -2949 & 0.007~  \\
(see Ref. \cite{Skripnikov:14a,Skripnikov:14b}) &   &  & & &  & &   \\  
 
\hline\hline
\end{tabular}

$^*$ The dipole moment is calculated with respect to Th nucleus.
\end{table*} 

We have applied the AIC theory described in the previous section to the case of ThF$^+$ and ThO.
To set the radial reference functions $\{\eta_{lj}^{f,g}\}$ given in~(\ref{eta}) we have used $7s$, $7p$ and $6d$ functions 
% (with appropriate $l,j,m$ numbers)
from calculation of the $7s^2 7p^1 6d^1$ configuration of Th and have evaluated the W-reduced density matrix $\Delta_{ljm, l'j'm'}$ defined by Eq.~(\ref{WRed}) from the molecular density matrices obtained within the CCSD approach. Note that $\{{\cal H}_{ljm}\}$ (\ref{eta}) coincide with the $7s$, $7p$ and $6d$ functions within some radius $R_c$ (here we set $R_c=0.25$ a.u.) and are zero outside the radius. For brevity we will designate the selected reference functions as $\widetilde{7s}_{1/2,1/2}$, etc. The operator of hyperfine interaction mixes the states with same parity and $m$;
%new
% the matrix element of the operator between 
%%
%$|ljm\rangle$ and $|l'jm\rangle$ 
%%
%is opposite by sign to the matrix element between 
%%
%%$lj-m$ and $l'j-m$ state.
%$|lj-m\rangle$ and $|l'j-m\rangle$ state.
%%
% Thus, the mean value of the operator is defined by the difference between the density matrix elements,
%$\Delta_{ljm, l'j'm} - \Delta_{lj-m, l'j'-m}$, and the hyperfine operator terms between the $|ljm\rangle$ and $|l'j'm\rangle$ states.
the diagonal matrix element of the operator for $|ljm\rangle$ state
is opposite by sign to the diagonal matrix element for $|lj-m\rangle$ state.
Thus, the diagonal contribution to the mean value of the operator is defined by the difference between the density matrix elements,
$\Delta_{ljm, ljm} - \Delta_{lj-m, lj-m}$, and the diagonal hyperfine operator terms for $|ljm\rangle$ states.
%end new
%
 For the most important elements of  W-reduced density matrix for the ThF$^+$ $^3\Delta_1$ state we have:
\begin{equation}
\begin{array}{l}                 
\Delta_{\widetilde{7s}_{1/2,1/2}, \widetilde{7s}_{1/2,1/2}} - \Delta_{\widetilde{7s}_{1/2,-1/2}, \widetilde{7s}_{1/2,-1/2}}=-0.99 \\
\Delta_{\widetilde{7p}_{1/2,1/2}, \widetilde{7p}_{1/2,1/2}} - \Delta_{\widetilde{7p}_{1/2,-1/2}, \widetilde{7p}_{1/2,-1/2}}=-0.47 \\
\Delta_{\widetilde{6d}_{3/2,3/2}, \widetilde{6d}_{3/2,3/2}} - \Delta_{\widetilde{6d}_{3/2,-3/2}, \widetilde{6d}_{3/2,-3/2}}=0.88 \\
\\ 
\Delta_{\widetilde{7s}_{1/2,1/2}, \widetilde{7p}_{1/2,1/2}} + \Delta_{\widetilde{7s}_{1/2,-1/2}, \widetilde{7p}_{1/2,-1/2}}=0.105 \\
\Delta_{\widetilde{7p}_{1/2,1/2}, \widetilde{7s}_{1/2,1/2}} + \Delta_{\widetilde{7p}_{1/2,-1/2}, \widetilde{7s}_{1/2,-1/2}}=0.105\ . \\
\label{RhoWElementsThF} 
\end{array}
\end{equation}

 For the most important elements of  W-reduced density matrix for the ThO $^3\Delta_1$ state we have:
\begin{equation}
\begin{array}{l}                 
\Delta_{\widetilde{7s}_{1/2,1/2}, \widetilde{7s}_{1/2,1/2}} - \Delta_{\widetilde{7s}_{1/2,-1/2}, \widetilde{7s}_{1/2,-1/2}}=-0.72 \\
\Delta_{\widetilde{7p}_{1/2,1/2}, \widetilde{7p}_{1/2,1/2}} - \Delta_{\widetilde{7p}_{1/2,-1/2}, \widetilde{7p}_{1/2,-1/2}}=-0.37 \\
\Delta_{\widetilde{6d}_{3/2,3/2}, \widetilde{6d}_{3/2,3/2}} - \Delta_{\widetilde{6d}_{3/2,-3/2}, \widetilde{6d}_{3/2,-3/2}}=0.67 \\
\\ 
\Delta_{\widetilde{7s}_{1/2,1/2}, \widetilde{7p}_{1/2,1/2}} + \Delta_{\widetilde{7s}_{1/2,-1/2}, \widetilde{7p}_{1/2,-1/2}}=0.238 \\
\Delta_{\widetilde{7p}_{1/2,1/2}, \widetilde{7s}_{1/2,1/2}} + \Delta_{\widetilde{7p}_{1/2,-1/2}, \widetilde{7s}_{1/2,-1/2}}=0.238 \\
\label{RhoWElementsThO} 
\end{array}
\end{equation}
Thus, in terms of the reference functions the effective configuration of unpaired electrons of Th in ThF$^+$ is $\widetilde{7s}^{1.0}\widetilde{7p}^{0.5}\widetilde{6d}^{0.9}$, while in ThO it is $\widetilde{7s}^{0.7}\widetilde{7p}^{0.4}\widetilde{6d}^{0.7}$.
The leading matrix element of the HFS operator in the basis of reference functions is between the $\widetilde{7s}$ functions. The ratio of effective occupancies of $\widetilde{7s}\approx 1.4$. This explains the ratio of the HFS constants given in table \ref{TResultsEeff}.

The operator of effective electric field (\ref{Wd}) mixes states of opposite parity with the same $m$; matrix element of the operator between $|ljm\rangle$ and $|l'jm\rangle$ has the same sign as the matrix element between $|lj-m\rangle$ and $|l'j-m\rangle$ state. Thus, the mean value of the operator is defined by combination of the types 
%new
%$\Delta_{ljm, l'j'm} + \Delta_{lj-m, l'j'-m}$ 
$\Delta_{ljm, l'jm} + \Delta_{lj-m, l'j-m}$
%end new
of the W-reduced density matrix $||\Delta_{ljm, l'j'm'}||$ and matrix elements of \Eeff\ operator between 
%new
%$|ljm\rangle$ and $|l'j'm\rangle$
$|ljm\rangle$ and $|l'jm\rangle$
%end new
 states. Most important of the combinations of $||\Delta_{ljm, l'j'm'}||$ matrix elements for ThF$^+$ and ThO $^3\Delta_1$ are given in Eqs.~(\ref{RhoWElementsThF}, \ref{RhoWElementsThO}). From the equations one can see that the W-reduced overlap population between the $\widetilde{7s}$ and $\widetilde{7p}$ functions in ThO is twice larger than that in ThF$^+$. This explains the appropriately larger \Eeff\ in ThO, see table \ref{TResultsEeff}.

Note that the W-reduced density matrix $||\Delta_{ljm, l'j'm'}||$ (and minimal number of core-property matrix elements over the reference W-reduced functions) can be considered as a pretty concise description of the effective atoms in compounds state which is appropriate for ``almost quantitative'' calculation of the mean values of core-properties under consideration.

%###################
\section{Conclusion}
%###################

The parameters \Eeff, $W_{T,P}$ and $W_M$ which are required to interpret experimental measurements on the $^3\Delta_1$ state of ThF$^+$ in terms of fundamental quantities are calculated. The value of \Eeff\ in ThF$^+$ was found to be notably smaller than that in ThO. A quantitative explanation is given. On the other hand, \Eeff(ThF$^+$) is 1.6 times bigger than the effective electric field in the HfF$^+$ cation \cite{Petrov:07a, Fleig:13} which is under preparation for electron electric dipole moment search \cite{Cossel:12}. 

In the present paper we have implemented the concept of atoms in compounds and applied it to calculate the W-reduced density matrix for description of the effective state of Th in ThF$^+$ and ThO.  This matrix contains ``sufficient information'' to evaluate such physically observable properties as hyperfine structure constant, etc. whereas the conventional density matrix is excessive here.

According to our preliminary study, the electronic spectrum of ThF$^+$ is more dense compared to the ThO molecule. 
Its accurate theoretical investigation requires inclusion of quadruple cluster amplitudes as shown in Ref.~\cite{Barker:2012} and is also found in our preliminary study of $^3\Delta_1$--$^1\Sigma$ transition energy.
We plan to investigate it in our future study of  ThF$^+$ elsewhere.

%%%%%%%%%%%%%%%%%%%%%%%%%%%%%%%%%%%%%%%%%%%%%%%%%%%%%%%%%%%%%%%%%%%%%%%%%%%%%%%
%%%%%%%%%%%%%%%%%%%%%%%%%%%%%%%%%%%%%%%%%%%%%%%%%%%%%%%%%%%%%%%%%%%%%%%%%%%%%%%

\section*{Acknowledgement}

This work is supported by the SPbU Fundamental Science Research grant from Federal budget No.~0.38.652.2013, RFBR Grant No.~13-02-01406. 
The AIC code was developed and calculations of W-reduced density matrix were performed with the support of the Russian Science Foundation grant (project No. 14-31-00022).
L.S.\ is also grateful to the grant of President of Russian Federation No.MK-5877.2014.2 and Dmitry Zimin ``Dynasty'' Foundation. The molecular calculations were partly performed on the Supercomputer ``Lomonosov''.

%\bibliographystyle{./bib/apsrev}
%\bibliography{bib/JournAbbr,bib/SkripnikovLib,bib/QCPNPI,bib/TitovLib,bib/Kaldor,bib/PetrovLib,bib/Kudashov}

\end{document}